\documentclass[aps,prl,twocolumn,superscriptaddress,showpacs]{revtex4}
\usepackage{bm}

\newcommand{\vct}[1]{{\bf #1}}
\newcommand{\gvct}[1]{\bm{#1}}
\newcommand{\ket}[1]{|#1\rangle}
\newcommand{\mtrix}[3]{\langle \,#1\,|#2|\,#3\,\rangle}
\newcommand{\fslash}[1]{\!\not\!#1}
\newcommand{\kf}{k_{\text{F}}}
\newcommand{\ekf}{E_{\text{F}}}
\newcommand{\kfp}{k_{\text{p}}}
\newcommand{\kfn}{k_{\text{n}}}
\newcommand{\vf}{v_{\text{F}}}
\newcommand{\mitg}{{\mit\Gamma}}
\newcommand{\thp}[1]{\theta^{\text{(p)}}_{\vct{#1}}}
\newcommand{\thn}[1]{\theta^{\text{(n)}}_{\vct{#1}}}
\newcommand{\ep}[1]{E_{\vct{#1}}}

\begin{document}

\title{The role of the Dirac sea in the quenching of Gamow-Teller strength}

\author{Haruki Kurasawa}
\affiliation{Department of Physics, Faculty of Science, Chiba University,
Chiba 263-8522, Japan}

\author{Toshio Suzuki}
\affiliation{Department of Applied Physics, Fukui University,
Fukui 910-8507, Japan\\
RIKEN, 2-1 Hirosawa, Wako-shi, Saitama 351-0198, Japan
}

\author{Nguyen Van Giai}
\affiliation{
Institut de Physique Nucl$\acute{e}$aire,
CNRS-IN2P3, 91406 Orsay Cedex, France
}

\begin{abstract}
The issue of Gamow-Teller sum rule in atomic nuclei is reexamined in the
framework of a relativistic description of nuclei. It is shown that the
total Gamow-Teller strength in the nucleon sector is reduced by about 12\%
as compared to the prediction of a non-relativistic description. This
reduction is due to the antinucleon degrees of freedom.
\end{abstract}


\maketitle

%%%%%%%%%%%%%%%%%%%%%%%%%%%%%%
The quenching of the observed Gamow-Teller (GT) strength in atomic nuclei is
a long standing issue both experimentally and theoretically. On the
experimental side the recent measurements\cite{s} have pinned down the
amount of strength reduction to about 10\% in medium-heavy nuclei.
Theoretically there are several possible sources of quenching ranging from
multiconfiguration spreading\cite{mult} to coupling to the $\Delta$-hole
sector\cite{delta}. The nuclear GT transitions are essential to many
important processes in particle and astrophysics related to neutrino-nucleus
interactions and therefore, further understanding of the GT quenching is
needed. In this work we point out a new quenching mechanism due to the
effects of the Dirac sea states.

We start from the model-independent 
Ikeda-Fujii-Fujita sum rule\cite{iff}
\begin{eqnarray}
\mtrix{}{Q_+ Q_-}{} - \mtrix{}{Q_- Q_+}{}
=2(N-Z)\,,\label{sr}
\end{eqnarray}
where 
\begin{eqnarray}
Q_\pm=\sum_i^A\left(\sigma_{y}\tau_{\pm}\right)_i~, \label{eq}
\end{eqnarray}
with the isospin operators $\tau_\pm=\left(\tau_x \pm
i\tau_y\right)/\sqrt{2}$. 
If we assume that there is no ground-state correlation,
\begin{eqnarray}
Q_+ \ket{\ } = 0\,,\label{s-}
\end{eqnarray}
and Eq.(\ref{sr}) becomes simply 
\begin{eqnarray}
\mtrix{}{Q_+ Q_-}{} = 2(N-Z).
\end{eqnarray}

In this work we study the GT sum rule
in a relativistic model where the nucleus is assumed to be composed of
Dirac particles bound in Lorentz scalar and vector potentials.
This  model has been extensively studied for the last 30 years,
and shown to describe very well both nuclear reactions and structure
phenomenologically\cite{sw}. 
We will show that about 12\% of
the sum rule value Eq.(\ref{sr}) is taken by nucleon-antinucleon states
with high excitation energy. As a result the strength of the giant
GT resonance is quenched by this amount. Moreover,
even if there is no ground-state correlations, Eq.(\ref{s-}) does
not hold, since there are nucleon-antinucleon excitations.

Let us start with the field:
\begin{eqnarray}
 \psi(\vct{x})&=&\int\!\frac{d^3p}{(2\pi)^{3/2}}\sum_{\alpha}
\left(
u_\alpha(\vct{p})\exp(i\vct{p}\cdot\vct{x})\,a_\alpha(\vct{p})
\right. \nonumber \\
& &
\left.
+\,v_\alpha(\vct{p})\exp(-i\vct{p}\cdot\vct{x})\,b^\dagger_\alpha(\vct{p})
\right)\,,
\end{eqnarray}
where the two terms describe the particles and the antiparticles, 
respectively. 
The suffix $\alpha$ denotes the spin and isospin quantum numbers as 
$u_\alpha(\vct{p})=u_\sigma(\vct{p})\,\ket{\tau}$\,, and the positive and 
negative energy spinors are written as
\begin{eqnarray}
u_\sigma(\vct{p})&=&\sqrt{\frac{\ep{p}+M^{\ast}}{2\ep{p}}}\left(
\begin{array}{cc}
1\\
\noalign{\vskip4pt} 
\displaystyle{
\frac{\gvct{\sigma}\cdot\vct{p}}{\ep{p}+M^{\ast}}
}
\end{array}
\right)\xi\,, \nonumber \\
\noalign{\vskip4pt}
v_\sigma(\vct{p})&=&
\sqrt{\frac{\ep{p}+M^{\ast}}{2\ep{p}}}
\left(
\begin{array}{cc}
\displaystyle{
\frac{\gvct{\sigma}\cdot\vct{p}}{\ep{p}+M^{\ast}}
}\\
\noalign{\vskip4pt}					   
1							      
\end{array}
\right)\xi\,
\end{eqnarray}
with $\ep{p}=\sqrt{M^{\ast2}+\vct{p}^2}$ and $\xi$ is the Pauli spinor.
The Lorentz vector potential does not appear explicitly in the
present discussions, while the Lorentz scalar potential does
in the nucleon effective mass $M^*$. Using the above field, 
an operator $F$ can be expressed as
\begin{eqnarray}
F=\int\!d^3x\,\overline{\psi}(\vct{x})\,f\,\psi(\vct{x})
\end{eqnarray}
where $f$ is considered to be some $4\times 4$ matrix with no
momentum-transfer dependence. From now on we will discuss nuclei
where $\kfn\ge\kfp$, $\kfn$ and $\kfp$ being the Fermi momentum of
neutrons and protons, respectively. Then, we have for the $\tau_-$
and $\tau_+$ excitations 
\begin{eqnarray}
F_-\ket{\ }&=&\sqrt{2}
 \int\!d^3p\,\sum_{\sigma\sigma'}
\Bigl(
\overline{u}_{\sigma}(\vct{p}) \mitg u_{\sigma'}(\vct{p})\,
a^\dagger_{\sigma \text{p}}(\vct{p})a_{\sigma'\text{n}}(\vct{p}) \Bigr.
\nonumber\\
& &
\left.
 +\,
\overline{u}_\sigma(\vct{p}) \mitg v_{\sigma'}(-\vct{p})\,
a^\dagger_{\sigma\text{p}}(\vct{p})
b^\dagger_{\sigma'\text{n}}(-\vct{p}) \right)\ket{\ }, \\
\noalign{\vskip4pt}
F_+\ket{\ }
&=&\sqrt{2}
\int\!d^3p\,\sum_{\sigma\sigma'}
\overline{u}_\sigma(\vct{p}) \mitg v_{\sigma'}(-\vct{p}) \nonumber \\
& &
\times \,a^\dagger_{\sigma\text{n}}(\vct{p})
b^\dagger_{\sigma'\text{p}}(-\vct{p})\,\ket{\ },
\end{eqnarray}
where $\mitg$ stands for the $4\times 4$ matrix other than the isospin
operator. Straightforward calculations yield 
\begin{eqnarray}
\mtrix{}{F_+F_-}{}
&=&
\frac{2V}{(2\pi)^3} \int\!d^3p\,\left(1-\thp{p}\right)
\left( \thn{p}\lambda_+ - \lambda_- \right)
\label{+}, \nonumber \\
\noalign{\vskip4pt} 
\mtrix{}{F_-F_+}{}
&=&
-\,\frac{2V}{(2\pi)^3} \int\!d^3p\,
\left(1-\thn{p}\right)\lambda_-, \label{-}
\end{eqnarray}
where we have used the fact that $\delta(0)=V/(2\pi)^3$,
$V=A\left(3\pi^2/2\kf^3\right)$ being the
volume of the system, and the abbreviations:
\begin{equation}
 \lambda_\pm
 = \text{Tr}\left( \mitg{\mit\Lambda}_+ \mitg{\mit\Lambda}_\pm \right)
 \label{lambda}
\end{equation}
with
\begin{eqnarray}
{\mit\Lambda}_+
&=&\sum_\sigma u_{\sigma}(\vct{p})\,\overline{u}_{\sigma}(\vct{p})
=\frac{\fslash{p}+M^\ast}{2\ep{p}}, \\ 
{\mit\Lambda}_-
&=&-\sum_\sigma v_{\sigma}(-\vct{p})\,\overline{v}_{\sigma}(-\vct{p})
=\frac{\fslash{\tilde{p}}+M^\ast}{2\ep{p}}
\end{eqnarray}
and $\tilde{p}^\mu=(-\ep{p},\vct{p})$. The step function is also
defined as $\theta_{\vct{p}}^{(i)}=\theta(k_i -|\vct{p}|)$ for
$i=\text{p}$ or n. In the case of the GT transition 
$\mitg=\gamma_5\gamma_y$, the traces of Eq.(\ref{lambda})
 are calculated as 
\begin{eqnarray}
\lambda_+
=2\frac{M^{\ast2}+p_y^2}{\ep{p}^2}\,,\quad
\lambda_-
=-\,2\frac{\vct{p}^2-p_y^2}{\ep{p}^2}.
\end{eqnarray}
Finally, we obtain the matrix elements
for the $\beta_-$ and $\beta_+$ GT transitions  
\begin{eqnarray}
\mtrix{}{F_+F_-}{} 
&=&
\frac{4V}{(2\pi)^3}\int\!\frac{d^3p}{\ep{p}^2}\,\left(1-\thp{p}\right)
\nonumber \\
& &
\times\left[
\thn{p}\left(M^{\ast2}+p_y^2\right)
+(\vct{p}^2-p_y^2) \right],\label{b+}\\
\noalign{\vskip4pt}
\mtrix{}{F_-F_+}{} 
&=&
\frac{4V}{(2\pi)^3}\int\!\frac{d^3p}{\ep{p}^2}\,
\left(1-\thn{p}\right)\left(\vct{p}^2-p_y^2 \right).\label{b-}
\end{eqnarray}
As seen from the above derivation, the second term in the square brackets
of Eq.(\ref{b+}) and the r.h.s. of Eq.(\ref{b-}) come from
the nucleon-antinucleon excitations, and each of them is divergent.
However, the difference between Eq.(\ref{b+}) and Eq.(\ref{b-})
gives the sum rule value as it should be, 
\begin{eqnarray}
\mtrix{}{F_+F_-}{} - \mtrix{}{F_-F_+}{} = 2(N-Z).\label{tsum}
\end{eqnarray}
If we neglect the antinucleons, then 
$F_+\ket{\ }=0$ while the first term in the square brackets
in Eq.(\ref{b+}) gives
\begin{eqnarray}
\mtrix{}{F_+F_-}{}_{\text{N}\text{N}}
&=&\frac{4V}{(2\pi)^3}\int\!\frac{d^3p}{\ep{p}^2}
\left(\thn{p}-\thp{p}\right)\left(M^{\ast2}+p_y^2\right)\nonumber\\
&=&\frac{A}{\kf^3}\left[\frac{\kfn^3}{3}+2M^{\ast2}\left(\kfn-M^{\ast}
\tan^{-1}\frac{\kfn}{M^\ast}\right)\right] \nonumber \\
\noalign{\vskip4pt}
& &
 -\,(\kfn \rightarrow \kfp).
\end{eqnarray}
Expanding the above equation up to first order in $(\kfn-\kfp)$, we
obtain
\begin{eqnarray}
\mtrix{}{F_+F_-}{}_{\text{N}\text{N}}
\approx
2\left(1-\frac{2}{3}v_{\text{F}}^2\right)\left(N-Z\right)~,\label{sum}
\end{eqnarray}
where $v_{\text{F}}$ is the Fermi velocity
\begin{eqnarray*}
v_{\text{F}}=\frac{\kf}{\ekf}\,,\quad
\ekf=\sqrt{M^{\ast2}+\kf^2}
\end{eqnarray*}
and we have assumed as usual that
\begin{equation}
(\kfn-\kfp)/\kf\approx 2(N-Z)/3A.\label{diff}
\end{equation}
Thus, the sum rule value in the nucleon space 
is quenched. In most of the relativistic
models\cite{cen} the nucleon effective mass is about $0.6M$, which
yields $\vf = 0.43$ for $\kf = 1.36\,\text{fm}^{-1}$. This 
corresponds to a quenching of 12\%.
If we use the free nucleon mass, 
the quenching is about 5\%.
The fraction of sum rule taken by nucleon-antinucleon excitations:
\begin{eqnarray}
[\mtrix{}{F_+F_-}{} - \mtrix{}{F_-F_+}{}]_{\text{N}\bar{\text{N}}}
\approx 2\,\frac23\vf^2\ (N-Z)
\end{eqnarray}
must be distributed over the higher excitation-energy region, but we need
the renormalization for more details.

The amount of quenching depends on the quasi-particle distribution,
since the commutation
relation taking only the nucleon degrees of freedom becomes 
\begin{eqnarray}
\left[\, F_+\,,\,F_-\,\right]
&=& 2\sum_{\sigma} \int\! d^3p\,
\left(1-\frac{\vct{p}^2-p_y^2}{\ep{p}^2}\right) \nonumber \\
& &
\times 
\Bigl(
  a^\dagger_{\sigma\text{n}}(\vct{p})a_{\sigma\text{n}}(\vct{p})
- a^\dagger_{\sigma\text{p}}(\vct{p})a_{\sigma\text{p}}(\vct{p})
\Bigr).
\end{eqnarray}
If quasi-particles have high-momentum components, the amount of 
quenching would be increased.
In the limit $\kf \rightarrow \infty$, the ratio of the GT
strength of nucleon excitations to that of antinucleon excitations is
1 to 2.

We note that in the case of the sum rule for the isobaric analogue
 state, which is similar to Eq.(\ref{tsum}), there is no contribution
 from the Dirac sea. This fact is shown in the same way as
for the GT sum rule. Taking ${\mit\Gamma}=\gamma_0$ in
 Eq.(\ref{lambda}),
 we obtain
$\lambda_+=2$ and $\lambda_-=0$, which lead to
\begin{eqnarray}
\mtrix{}{F_+F_-}{}=2(N-Z)\,,\quad
\mtrix{}{F_-F_+}{}=0
\end{eqnarray}
instead of Eqs.(\ref{b+}) and (\ref{b-}). 

Since a part of the GT strength is taken by the
nucleon-antinucleon states, we investigate the effects of the Pauli blocking
terms on the strength of the giant GT resonance in RPA.
It has been shown that the Pauli blocking terms play an important
role for some observables in the relativistic model\cite{dirac}. 
We assume that the RPA correlations are induced through
the Lagrangian\cite{ksg}:
\begin{eqnarray*}
 {\cal L}
= \frac{g_5}{2}\,
 \overline{\psi}\mitg^\mu_i\psi\,
 \overline{\psi}\mitg_{\mu i}\psi,\quad
\mitg^\mu_i=\gamma_5\gamma^\mu\tau_i\,,\quad
g_5=g'\frac{f_\pi^2}{m_\pi^2}, 
\end{eqnarray*}
where $g'$ stands for the Landau-Migdal parameter. Although the way
to introduce $g'$ in the relativistic model is model-dependent\cite{ksg},
we take this form which reproduces the non-relativistic result for
the excitation energy of the giant GT resonance, as will be
shown later. Then, the RPA correlation function 
${\mit\Pi}_{\text{RPA}}$ for the $\tau_-$ excitations is written in terms of
the mean field one ${\mit\Pi}$, 
\begin{eqnarray}
{\mit\Pi}_{\text{RPA}}(\mitg^\mu_+\,, \mitg^{\nu}_ -)
= [U^{-1}]^{\mu\lambda}{\mit\Pi}(\mitg_{\lambda+}\,, \mitg^{\nu}_ -),
\end{eqnarray}
where $U$ denotes the dimesic function: 
\begin{eqnarray}
U^{\mu\nu}=
g^{\mu\nu}-\chi_5{\mit\Pi}(\mitg^\mu_+\,, \mitg^\nu_-\,)
\end{eqnarray}
with $\chi_5=g_5/(2\pi)^3$. The general form of the mean field
correlation function including the Pauli blocking terms
can be found in Refs.\cite{chin,445}.
We follow the usual procedure of taking into
account all the terms containing the density-dependent part
$G_{{\rm D}}$ of the single-particle propagator $G$, but neglecting terms like
$\mitg_\alpha G_{\text{F}}\mitg_\beta G_{\text{F}}$ ($G_{\text{F}}$ is the
Feynman part of $G$) 
which may be divergent\cite{chin, 445} and should be treated by a
renormalization procedure. 
For ${\mit\Pi}( \mitg^{\mu}_+ , \mitg^{\nu}_-\,)$ one obtains 
\begin{widetext}
\begin{eqnarray}
{\mit\Pi}( \mitg^{\mu}_+ , \mitg^{\nu}_-\,)
&=&
\int\!\!d^4p\,\frac{\delta(p_0-E_{\vct{p}} )}{\ep{p}}
\left(
\frac{t^{\mu\nu}(p,q)}{(p+q)^2-M^{\ast2}+i\varepsilon}
\,\thn{p}
+ \frac{t^{\mu\nu}(p,-q)}{(p-q)^2-M^{\ast2}+i\varepsilon}
\,\thp{p}
\right) \nonumber\\
 \noalign{\vskip4pt} 
& &
+\, i\pi
\int\!\!d^4p\,\frac{\delta(p_0-\ep{p} )\,
\delta(p_0+q_0-E_{\vct{p}+\vct{q}} )}{\ep{p} E_{\vct{p}+\vct{q}} }
t^{\mu\nu}(p,q)\,\thn{p}\theta^{\text{(p)}}_{\vct{p}+\vct{q}},
\label{mf}
\end{eqnarray}
\end{widetext}
where $t^{\mu\nu}(p,q)$ is 
\begin{eqnarray}
t^{\mu\nu}(p,q)
& = & 4\Bigl(
  g^{\mu\nu}\left(M^{\ast2}+p^2+p\!\cdot\!q \right) \nonumber \\
& - & 2p^\mu p^\nu -p^\mu q^\nu-p^\nu q^\mu  
\Bigr).
\end{eqnarray}
The above equations show that the dimesic function can be separated
into transverse and longitudinal parts. In the case of $\vct{q}=0$,
the determinant can be written as 
\begin{eqnarray}
\text{det}\ U =-\left(D_{\text{T}}\right)^3D_{\text{L}},
\end{eqnarray}
where we have defined
\begin{eqnarray}
D_{\text{T}}&=&1+\chi_5\,{\mit\Pi}(\mitg^{2}_+\,, \mitg^{2}_-\,),\\
\noalign{\vskip4pt}
D_{\text{L}}
&=&1-\chi_5 {\mit\Pi}(\mitg^{0}_+\,, \mitg^{0}_-\,)
\label{tl}\,.
\end{eqnarray} 

The eigenvalues of the excitation energies are obtained from the real
part of the dimesic function, 
\[
 \text{det\ Re}\ U=0.
\]
The longitudinal part has no solution, while the transverse part yields
for $q_0\ll M^{\ast}$ with Eq.({\ref{diff}})\cite{ksg}
\begin{eqnarray}
\omega_{\text{GT}}=\frac{1-2\vf^2/3}{1-g_5P(\kf)}\,
     g_5\frac{8\kf^3}{3\pi^2}\frac{N-Z}{2A},\label{gt}  
\end{eqnarray}
where $P(\kf)$ stems from the Pauli blocking terms,
\begin{eqnarray}
P(\kf) 
&=&\frac{4}{3\pi^2}\ekf^2
\left(
\frac32\vf-\vf^3-\frac34\left(1-\vf^2\right)\log\frac{1+\vf}{1-\vf}
\right)\nonumber \\
&=&
\frac{4}{15\pi^2}\kf^2\vf^3\left( 1+\frac37\vf^2+\cdots\right)\label{p}.
\end{eqnarray}
In the non-relativistic limit $\vf^2 \ll 1$, the above equation gives
\begin{eqnarray}
\omega_{\text{GT}} = g_5\frac{8\kf^3}{3\pi^2}\frac{N-Z}{2A},
\end{eqnarray}
which is just  
the excitation energy of the giant GT resonance
obtained in the non-relativistic model\cite{ts}. If we use the
value $g'$= 0.6\cite{dh}, 
the Pauli blocking effect on the
excitation energy is less than 0.5\%.

The strength of the giant GT resonance is given by the imaginary
part of the RPA correlation function.
Writing ${\mit\Pi}(q_0)={\mit\Pi}( \mitg^2_+ , \mitg^2_-\,)$,
we expand the transverse dimesic function $D_{\text{T}}(q_0)$ at 
$q_0=\omega_{\text{GT}}$,
\begin{eqnarray}
D_{\text{T}}(q_0)=\alpha_{\text{T}}\left(q_0-\omega_{\text{GT}}\right)
+\cdots\,,\quad
\alpha_{\text{T}}
=\left.\frac{dD_{\text{T}}}{dq_0}\right|_{q_0=\omega_{\text{GT}}}\, .
\end{eqnarray}
Then, the RPA correlation function is 
\begin{eqnarray}
{\mit\Pi}_{\text{RPA}}(q_0)
=\frac{1}{\alpha_{\text{T}}}
\frac{{\mit\Pi}(q_0)}{q_0-\omega_{\text{GT}}+i\varepsilon}, 
\end{eqnarray}
which gives the imaginary part
\begin{eqnarray}
 \text{Im}\,{\mit\Pi}_{\text{RPA}}(q_0)
=\frac{\pi}{\chi_5\alpha_{\text{T}}}
\delta(q_0-\omega_{\text{GT}}).
\end{eqnarray}
The relationship of the imaginary part to the strength\cite{445},
together with Eq.(\ref{gt}), provides us with the strength of the giant
GT resonance:
\begin{eqnarray}
S_{\text{GT}}=
\frac{2}{\left( 1-g_5P(\kf) \right)^2}
\left(1-\frac23\vf^2\right)\left(N-Z\right).
\end{eqnarray}
The above result shows that $S_{\text{GT}}$ is almost
equal to the one in the nucleon space given in Eq.(\ref{sum}),
since $P(\kf)$ is negligible. Thus, the Pauli blocking terms yield
almost no effect on the strength of the giant GT resonance. 
When we neglect $P(\kf)$, $S_{\text{GT}}$ is independent of $g_5$
and it exhausts the sum value in the nucleon space.

It should be noted that the Pauli blocking terms change the sum
value of the unperturbed states. The imaginary part of the mean
field correlation function is given by Eq.(\ref{mf}) at $\vct{q}=0$:
\begin{eqnarray}
& &
\text{Im}\, {\mit\Pi}( \mitg^2_+ , \mitg^2_-\,)\nonumber  \\
&=&
-\,\pi
\int\!\!d^4p\,\frac{\delta(p_0-\ep{p} )}{\ep{p}}
\Bigl[\,
 t^{22}(p,q)\,\delta( q_0^2+2p_0q_0 )\,\thn{p}
 \Bigr. \nonumber \\
& &
+\,t^{22}(p,-q)\,\delta( q_0^2-2p_0q_0 )\,\thp{p} \nonumber \\
& &
-\,\frac{\delta(q_0)}{\ep{p}}
\,t^{22}(p,q)\,\thn{p}\,\thp{p}
\,\Bigr
].
\end{eqnarray}
When we employ the relation:
\begin{eqnarray}
 \delta( q_0^2\pm 2p_0q_0)=\frac{1}{2p_0}\,\left(\delta(q_0)
+\delta(q_0\pm 2p_0)\right)\,,
\end{eqnarray}
the first and second $\delta$-functions of the r.h.s. 
correspond to the sum values from the nucleon space and 
from the Pauli blocking N$\bar{\text{N}}$-excitations, respectively. 
The latter yields the sum value for the $\beta_-$ transition:
\begin{eqnarray}
S_{\text{N}\bar{\text{N}}}=-\frac{3A}{4\pi\kf^3}\int\! d^3p\,
 \frac{\vct{p}^2-p_y^2}{\ep{p}^2}\thp{p},
\end{eqnarray}
which is negative, and equal to the density-dependent part
from N$\bar{\text{N}}$-excitations in Eq.(\ref{b+}).

In conclusion, we have shown that, in a relativistic description the total
GT strength in the nucleon sector is quenched by about 12\% 
owing to the antinucleon degrees of
freedom. The effects on the giant GT resonance energy are negligible. Of
course, the strength in the nucleon sector can be further spread out to the
tail of the giant resonance by configuration mixing effects similarly to
the non-relativistic picture. It has been discussed 
for a long time that the GT strength may
be quenched because of the coupling of the particle-hole states with
$\Delta$-hole states\cite{dh}. The present discussions predict
another source of the quenching. The recent experiments have observed
about 90\% of the classical sum rule value\cite{s}. One should
study in more detail how $\Delta$-hole and
N$\bar{\text{N}}$ states contribute to the observed quenching. 
It would be interesting to extend the present investigation to include the
diverging $\mitg_\alpha G_{\text{F}}\mitg_\beta G_{\text{F}}$ terms by a
renormalization procedure for a consistent treatment in the future.

Finally, we note that the Coulomb sum values are also strongly
quenched by the antinucleon degrees of freedom\cite{cos}. This fact
is consistent with recent analysis of experimental data\cite{m}.
Sum rules require the complete set of nuclear wave functions.
This implies generally that antinucleon degrees of freedom have a
possibility to modify the sum rule values as compared to non-relativistic
models.

\acknowledgments

The authors would like to thank Professors G. F. Bertsch
and Z. Y. Ma for useful discussions.
One of them(T. S.) appreciates a kind
hospitality of the Theory group of IPN at Orsay.

\end{document}